\documentclass[10pt, twocolumn]{article}
\usepackage{latex8}
\usepackage{algorithmic}
\usepackage{algorithm}
\usepackage{times}
\usepackage{latexsym}
\usepackage{setspace}
\usepackage[hmargin={0.95in,0.95in},vmargin={0.95in,0.945in}]{geometry}
\usepackage{graphicx}
\usepackage{subfigure}
\usepackage{balance}

\title{Decentralized Management of Bi-modal Network Resources \\ 
in a Distributed Stream Processing Platform }

\author{
\textit{Shah Asaduzzaman} \\
School of Information Tech.\ and Engg. \\
University of Ottawa\\
Ottawa, ON, K1N 6N5, Canada \\
Email: \texttt{asad@site.uottawa.ca}
\and 
\textit{Muthucumaru Maheswaran} \\
School of Computer Science \\
McGill University \\
Montreal, QC H3A 2A7, Canada \\
Email: \texttt{maheswar@cs.mcgill.ca}
}

\date{}

\begin{document}

%********************************************************************

\pagestyle{empty}
\maketitle
\thispagestyle{empty}

\begin{abstract}
  This paper presents resource management techniques for allocating
  communication and computational resources in a distributed stream
  processing platform. The platform is designed to exploit the synergy
  of two classes of network connections -- dedicated and
  opportunistic. Previous studies we conducted have demonstrated the
  benefits of such {\em bi-modal} resource organization that combines
  small pools of dedicated computers with a very large pool of
  opportunistic computing capacities of idle computers to serve high
  throughput computing applications. This paper extends the idea of
  bi-modal resource organization into the management of communication
  resources. Since distributed stream processing applications demand
  large volume of data transmission between processing sites at a
  consistent rate, adequate control over the network resources is
  important to assure a steady flow of processing. The system model
  used in this paper is a platform where stream processing servers at
  distributed sites are interconnected with a combination of dedicated
  and opportunistic communication links. Two pertinent resource
  allocation problems are analyzed in details and solved using
  decentralized algorithms. One is mapping of the processing and the
  communication tasks of the stream processing workload on the
  processing and the communication resources of the platform. The
  other is the dynamic re-allocation of the communication links due to
  the variations in the capacity of the opportunistic communication
  links. Overall optimization goal of the allocations is higher task
  throughput and better utilization of the expensive dedicated links
  without deviating much from the timely completion of the tasks. The
  algorithms are evaluated through extensive simulation with a model
  based on realistic observations. The results demonstrate that the
  algorithms are able to exploit the synergy of bi-modal communication
  links towards achieving the optimization goals.
\end{abstract}

%\doublespace

%Start--------------------------------------------------------------------
\section{Introduction}
\label{sec:intro}
Many applications on the Internet are creating, manipulating, and
consuming data at an astonishing rate. Data stream processing is one
such class of applications where data is streamed through a network of
servers that operate on the data as they pass through them~\cite{
  Repantis2009, Benoit2009, Hwang2008, Seshadri2007, Gates2004,
  Pietzuch2008, Pietzuch2006}. Depending on the application, data
streams can have complex topologies with multiple sources or multiple
sinks. Examples of data stream processing tasks are found in many
areas including distributed databases, sensor networks, and multimedia
computing. Some examples include: (i) multimedia streams of real-time
events that are transcoded into different formats~\cite{Gu2006}, (ii)
insertion of information tickers into multimedia
streams~\cite{Carney2002}, (iii) real-time analysis of network
monitoring data streams for malicious activity
detection~\cite{Gigascope2003}, and (iv) function computation over
data feeds obtained from sensor networks~\cite{Seshadri2007}.

One of the salient characteristics of this class of applications is
the simultaneous demand for high-throughput computing and
communication resources~\cite{Kalogeraki2007}. Huge volume of data
generated at high rates need to be processed within real-time
constraints. Moreover, various operations on these data streams are
provided by different servers at distributed geographic
locations~\cite{Nahrstedt2006}. All these factors demand a scalable
and adaptive architecture for distributed stream processing platform,
where fine-grained control over processing and network resources is
possible.

Earlier works on stream processing engines~\cite{Aurora2003,
  STREAM2003} resorted to centralized single-server or server-cluster
based solutions where tighter control over available resources is
possible. With the possibility of different processing services or
operations being provided by different providers, need for distributed
stream processing platform arose. Several architectures have been
proposed to support such distributed processing of
streams~\cite{Kalogeraki2007, Schwan2005, Nahrstedt2006,
  Karamcheti2004}. Due to the stringent rate-requirement for
processing and transmission of data, most researchers have assumed a
central resource controller that can gather the availability status of
all resources and map the requested tasks on them. However, with the
advent of a diverse range of stream processing services, it is
important to allow autonomous providers of services to collaborate and
share their resources. Thus it is important to develop decentralized
resource allocation schemes, where control is available over local
resources only.

While it is feasible to have dedicated server resources and precisely
allocate them for processing tasks, dedicated networks over wide-area
installations remain costly. Although it is possible to propagate the
data streams through the distributed servers using the Internet, the
lack of adequate control over end-to-end bandwidth on the Internet and
the stringent rate requirements of the stream processing applications
demand some dedicated network resources. In fact, recent advances in
optical network technologies such as user-controlled light
path~\cite{Bochmann2003, Boutaba2008} open the possibility of
on-demand provisioning of end-to-end optical links with total control
of the available bandwidth is exposed to the user application.

In this paper, we explore a novel approach where a combination of
dedicated and opportunistic communication links is used to
interconnect the servers. The main focus of this paper is to explore
how such a hybrid (denoted as {\em bi-modal} in this paper) network
can be best used for data stream processing tasks. The hypothesis that
drives this work is that the combination has a synergistic effect that
allows better utilization of the dedicated resources, and yields
higher return on investment. We devised distributed algorithms for
allocation of these hybrid resources to demonstrate the viability of
this synergy hypothesis.

Multiple global objectives such as higher task throughput, lower
violation of SLA and higher utilization of dedicated resources make
the resource management a complex task, especially when allocation
decisions are to be taken solely based on the local information
available on the server nodes. We divided the overall resource
management process into two steps -- initially individual tasks are
assigned node and link resources through a distributed mapping
algorithm. Based on actual resource availability, link resources are
then periodically re-allocated locally among competing tasks towards
the global optimization objectives.

This paper extends some of our previous works~\cite{jpdc07,JPDC06} on
bi-modal compute platforms where static small pool of dedicated
compute-servers was combined with a large number of opportunistically
harvested cheap processing elements to increase work throughput and
utilization of dedicated resources. Using data stream processing tasks
as a concrete example, this paper demonstrates the benefit of using
bi-modal network infrastructures for communication-intensive
applications. In particular, this paper makes the following
contributions to this important resource management problem:
\begin{itemize}
\item 
Show that the bi-modality of the network helps to improve the
utilization of dedicated resources such as servers and network links.
\item
Show that the bi-modal organization allows the platform to admit
significantly larger workload and yield significantly higher
throughput without deviating much from the service contracts.
\item 
Show the importance of adaptive scheduling to cope with the
variability in the capacity of the opportunistic network.
\end{itemize}

In Section~\ref{sec:model} we present the system model for the
distributed stream processing platform and assert the necessary
assumptions. Section~\ref{sec:problem} introduces and characterizes
the two resource management problems pertinent to the platform -- the
problem of mapping the tasks to the resources and the problem of
periodically re-allocating the resources to adapt with the
ever-changing behavior of the opportunistic resources.
Section~\ref{sec:mapping} explains our proposed solution to the
mapping problem. Section~\ref{sec:dynasched} explains the algorithm
for periodic re-allocation of the communication resources.  The
algorithms presented in both the Section~\ref{sec:mapping} and
Section~\ref{sec:dynasched} are local algorithms engineered to
gradually achieve some global optimization objectives such as high
throughput and resource utilization. In Section~\ref{sec:results}, we
evaluate through extensive simulations the extent to which these
global objectives are achieved by the algorithms. We then conclude
with a discussion of related literature in Section~\ref{sec:related}.

\section{System Model and Assumptions}
\label{sec:model}

\subsection{System Model}
In a {\em stream processing task}, the data stream originating from a
{\em data-source} node, progresses through several steps of processing,
termed as {\em service components} (or {\em service} in short), before
being delivered to the {\em data-delivery} node. For example, in video
streaming, the service components may be encoding of video, embedding
some real time tickers and transcoding the video into different
formats. Although, in very general terms, the data-flow topology could
be arbitrary graphs, in this paper, we restrict our study to
simple path topologies.

The distributed stream processing platform consists of several
autonomous server nodes that serve the service components. A single
server may serve multiple services and a service may be available at
multiple servers. Several pairs of servers establish dedicated
point-to-point links between them to have the flow of the data streams
at a controlled rate. Each server is also connected to the public
Internet and end-to-end TCP connection can be established between any
pair of servers through the Internet. However, with the Internet,
end-to-end bandwidth of the TCP connections cannot be allocated and
the flow rate cannot be controlled. These connections are thus treated
as {\em opportunistic} resources. Both the dedicated and opportunistic
links are assumed to be bi-directional and of symmetric capacity, for
both data-transport and control messaging purposes. The assumption on
the bi-directionality of data-transport is not absolutely necessary
for such platforms, the assumption is rather made for the convenience
of discourse.

The platform is modeled as an asynchronous message passing distributed
system, where there is no centralized controller to coordinate the
resources. The servers have knowledge of and can precisely allocate
the local resources only, i.e. the processing capacity of the node and
the bandwidth of the outgoing communication links. However, the
servers comply with the global protocol and respond to a predefined
set of messages in a predefined way. The objective of the global
protocol is to ensure adequate resources for each individual task for
its seamless progress, and to maximize the global work
throughput. Other factors such as balancing the load among different
servers and maximizing the utilization of dedicated resources are also
considered. Design and evaluation of the protocol constitute the
remaining sections of the paper.

\begin{figure*}[htbp]
  \begin{minipage}[t]{0.58\linewidth}
    \centering
    \includegraphics{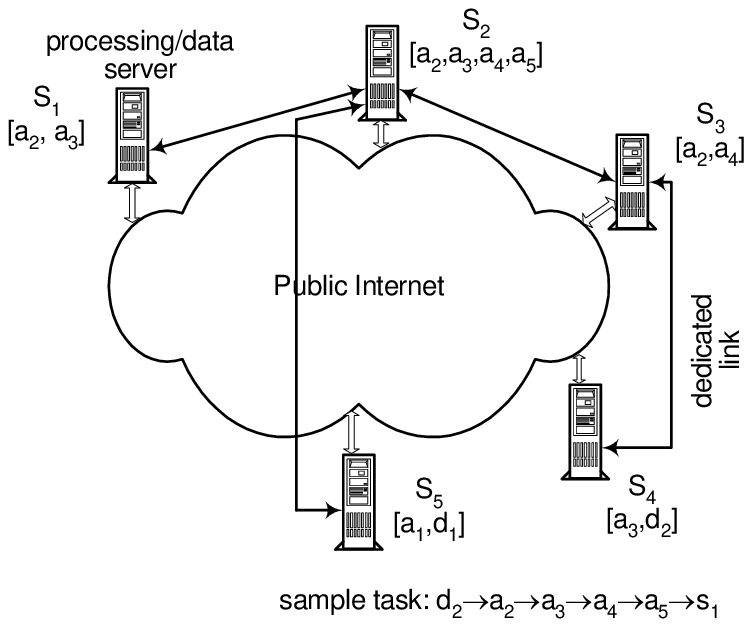}
    \caption{Stream processing platform}
   \label{fig:arch_stream_arch}
   \end{minipage}
   \hspace{0.5cm}
   \begin{minipage}[t]{0.38\linewidth}
     \centering  
     \includegraphics{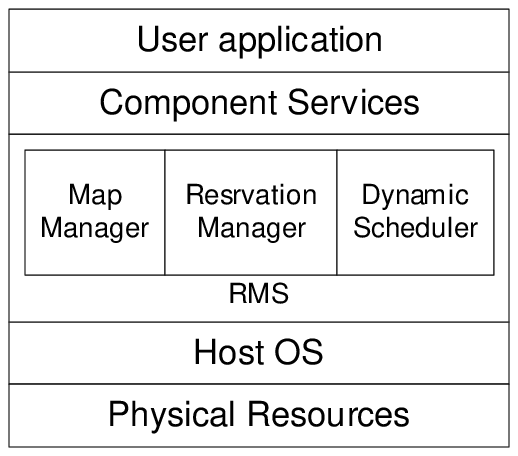}
     \caption{Layered architecture}
     \label{fig:arch}
   \end{minipage}
\end{figure*}

Figure~\ref{fig:arch_stream_arch} illustrates a scenario of a stream
processing platform containing five servers. The example stream
processing task shown in the figure requests a data stream from data
source $d_2$ to be processed through services $a_2$, $a_3$, $a_4$ and
$a_5$, and to be delivered to $S_1$. This task
may be served by the servers $S_4$ (serving $d_2$), $S_3$ (serving
$a_2$), $S_2$ (serving $a_3$ and $a_4$). Either dedicated link or
public network may be used to transmit the data stream between any two
consecutive servers.

For convenience, the resource allocation process is divided into two
phases. First, individual tasks with multiple service components are
mapped on the processing servers fulfilling the processing and
transport capacity requirements. A cost function is used to select the
best among multiple feasible maps. The second phase re-allocates the
link bandwidths among competing tasks, after the tasks start execution
based on the initial allocation. This is necessary because of the
variability of data rate in the end-to-end TCP connections on the
Internet. Both the re-allocation phases and initial allocation are
driven by the same global optimization goal, namely maximization of
global throughput and resource utilization, subject to fulfillment of
individual task requirements.

\subsection{Architecture}
\label{subsec:model_arch}
The stream processing platform can be viewed to be composed of the
layers showed in Figure~\ref{fig:arch}, with user applications at the
top. The applications are composed of data sources and several service
components hosted by different servers. Therefore, the service
components constitute the next layer. At the bottom layer, the
resource management system (RMS) of the platform manages the available
server and network resources to allow seamless execution of the
service components. The main focus of this paper is to design and
analyze the algorithms for various functionalities of the RMS
layer. The RMS is responsible for mapping of the task requests on
available resources and dynamically adapting the resource allocations
in response to various loading conditions. The two components of RMS
cooperate to achieve these functionalities. A detailed discussion on
the RMS is presented in Section~\ref{sec:problem}. RMS uses the local
operating system API to control the underlying resources. Hence host
OS and physical resources lie at the bottom of the layered
architecture.

\subsection{Task Specification}
\label{subsec:model_task}
The specification of the stream processing task includes the ordered
sequence of service components, the data source node, the data
delivery node and the desired rate of data delivery. We assume a rate
based model to specify resource requirement for each service
component. For any service, both the output data rate and the CPU
requirement are proportional to the input data rate, and are specified
by two factors -- the {\em bandwidth shrinkage factor} and the {\em
  CPU usage factor}, respectively. We assume that these two factors
for any service component is known globally. Thus any node receiving
the task specification can compute the CPU and input/output data rate
requirements for each service component. This rate based model is
similar to the ones used by Kichkaylo et al.~\cite{Karamcheti2004} and
Drougas et al.~\cite{Kalogeraki2007}.

The task specification is a {\em service level agreement} (SLA)
between the user and the platform. On receiving the request for
resource for a task, the platform attempts to allocate necessary
resources. The platform may be unsuccessful to allocate all necessary
resources due to the loading condition of the platform, and the task
may be rejected as a result. Once the task is accepted after
successful resource allocation, it is responsibility of the platform
to meet the constraints specified in the SLA.

\subsection{Pricing and Revenue Flow}
\label{subsec:model_pricing}
We assume a rate based pricing for the services. The task
specification includes a price per byte of data delivered. This price
quote is directly translated to apportioned revenue for each of the
service components, using the CPU usage and bandwidth shrinkage
factors. The server that serves a service component receives revenue
for each byte processed at this apportioned rate. In some cases, some
server may need to forward the data without any processing, due to the
particular task-to-resource mapping chosen. We assume there is a
universally defined price charged by any server for per byte of data
forwarding. Because the data forwarding path for service $i$ to
service $i+1$ is chosen by the server of service $i$, it is assumed
that any forwarding price incurred before reaching the server serving
the $(i+1)$th service is paid by the previous server.

 %system model and assumptions
\section{Decentralized Management of Server and Network Resources}
\label{sec:problem}
A resource management engine (denoted as RMS agent) runs in each
server that implements the protocols for coordinated allocation of
network and CPU resources.  The resource management process is divided
into two phases -- initial mapping of individual tasks and dynamic
re-allocation of the resources among competing tasks. Accordingly,
each RMS agent has two modules -- a map manager and a dynamic
scheduler. This section defines the two problems in details and
illustrates the global picture that integrates these two phases for
global resource management objectives. The following two sections
discusses the possible solutions to these problems.

A user of the distributed platform uses one of the server nodes as a
portal to launch her stream processing task. The task specification
submitted to the portal contains the address of data stream source and
an ordered list of the service components that should process the data
stream. By default the delivery point (destination) of the stream is
the user's portal node, but any other node can be specified as well.
The specification also includes the required rate of data delivery,
time window for monitoring the rate and pricing for each byte of data
delivered. The parameters such as data rate and pricing may be
negotiated between the user and the portal through an automated SLA
negotiation protocol, details of which is out of the scope of this
paper.

\begin{figure*}[htbp]
\begin{minipage}[b]{0.48\linewidth}
\centering
\includegraphics{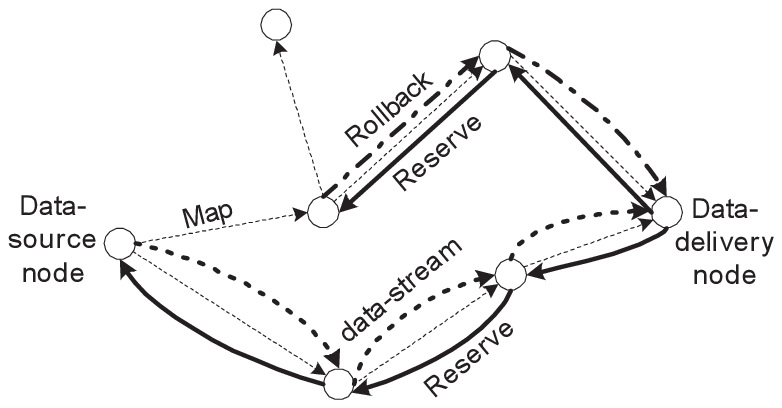}
\caption{Mapping, reservation and rollback}
\label{fig:map_reserve}
\end{minipage}
\hspace{0.5cm}
\begin{minipage}[b]{0.48\linewidth}
\centering
\includegraphics{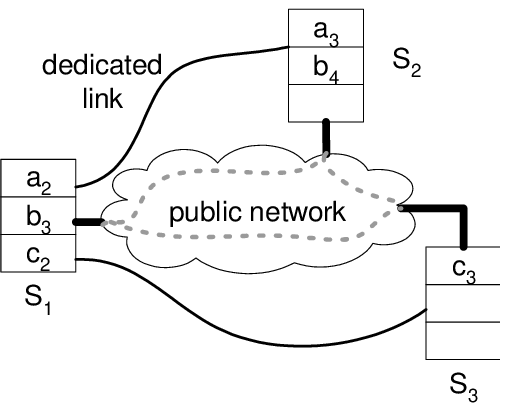}
\caption{Dynamic scheduling of link resources done by server $S_1$ on
  three competing task segments $a_2$-$a_3$, $b_3$-$b_4$, $c_2$-$c_3$}
\label{fig:dynasched}
\end{minipage}
\hspace{0.5cm}
\end{figure*}

After receiving the specification from user, the portal node initiates
the mapping process by sending a map message with the initial mapping
and the requirement specification to the data-source node.  Through
message passing among the map managers in different server nodes, the
distributed mapping algorithm results in a set of feasible maps at the
map manager of the data-delivery node.  Each of the maps defines a
path from the data source node to the delivery node through the server
nodes that serve necessary service components. The best among the
available feasible maps according to a certain cost metric is
selected. We assume that the cost metric is additive and the cost is
incurred at every node and link used by the task.

A reservation probe is then sent from the data-delivery node to the
data-source node along the path found in the selected map. The RMS
agent at each server node along the path tries to allocate the server
and link resources prescribed by the map. Because the mapping process
for multiple tasks may be ongoing concurrently, it is possible that
the required resource is no longer available. In such case the
allocation fails, the probe is rolled back and the next feasible map
is probed by the data-delivery node.  The streaming and the execution
of the stream processing task begins once a successful probe reaches
the data-source node at the other end. The message flow of mapping and
reservation is illustrated in Figure~\ref{fig:map_reserve}.

\subsection{The Mapping Problem}
\label{subsec:prob_mapping}
Abstracting away the details of the two classes of communication links
and different types of service, the mapping of a stream processing
task on the network of servers can be described as a problem of
constrained mapping of a weighted directed path on a weighted
undirected graph.

The network of servers can be defined as a graph $G_R = (V_R, E_R)$.
Each vertex $v_R \in V_R$, denotes a server that has an available
computational capacity $C_{av}(v_R)$. Each edge $e_R \in E_R$ denotes
a data transport link with an available bandwidth $B_{av}(e_R)$. Each
edge $e_J$ also has an associated additive cost $W(e_R)$. The stream
processing task can be defined as a path $P_J = (V_J, E_J)$, $V_J =
{v_0=s_J, v_1, v_2, ..., v_m = t_J}$ and $E_J = \{e_i = (v_i, v_{i+1})
| 0 \leq i < m \}$. Each vertex $v_i, 0 \leq i \leq m$ of the stream
processing task has a computational capacity requirement
$C_{req}(v_i)$, and each edge $e_i = (v_i, v_{i+1}), 0 \leq i < m$ has
a bandwidth requirement $B_{req}(e_i)$.

The problem is to find mappings $M_v : V_J \rightarrow V_R$ and $M_e:
E_J \rightarrow P_R$, where $P_R$ is the set of all possible paths in
the resource graphs, including zero length paths. The second mapping
$M_e$ is needed because a server node can act as forwarding nodes
and thus, each edge in $E_J$ can potentially be mapped on a multi-hop
path $p_R$ in $G_R$. Also, multiple vertices from $V_J$ can be mapped
on a single vertex of $V_R$, which essentially maps edges from $E_J$
on zero length paths, i.e. $(v,v)$ paths with infinite bandwidth and
zero cost. Again, it is allowed that for two different edges, $e_1,
e_2 \in E_J$, the mapped paths $p_1 = M_e(e_1)$ and $p_2 = M_e(e_2)$
have some common edges. The mapping of the source node and the sink
node is already given: $M(s_J) = s_R | s_R \in V_R$ and $M(t_J) = t_R
| t_R \in V_R$.

The mapping has to fulfill the following constraints on processing
capacity and bandwidth --

\begin{eqnarray}
\forall{v_R \in M(v_J)}, \sum_{\{ v_J | v_J \in V_J, M(v_J) = v_R\}} {C_{req}(v_J)} 
     &\leq & C_{av}(v_R) \nonumber
\end{eqnarray} 
$$\forall {e_J = (u,v) \in E_J}, B(e_J) \leq min[ B(e_R), e_R \in M_e(e_J)]$$ 

The constraints define the decision problem -- ``Is there any $M$ and
$M_e$ that satisfies the constraints?''. This problem can be proved to
be NP-complete by transformation to the longest path
problem~\cite{GareyJohnsonBook}. The details of the proof can be found
at~\cite{HPCS2007,AsadPhDThesis}. When the result of the decision
problem is true, there can be multiple feasible mappings that
satisfies the constraints. To choose a single mapping among the
feasible ones, we can formulate a corresponding optimization problem,
where each edge $e_R \in E_R$ in the resource graph has an additive
cost $W(v_R)$. The objective would be to find the feasible mapping
that minimizes the total cost $W = \sum{W(p_R)} | p_R \in M_e(u,v)
\forall{u,v} \in V_J$. Cost $W(p_R)$ of a path $p_R$ is the sum of the
costs $W(e_R)$ of all edges $e_R$ in $p_R$.

\begin{figure*}[htbp]
\begin{minipage}{0.48\linewidth}[b]
\centering
\includegraphics{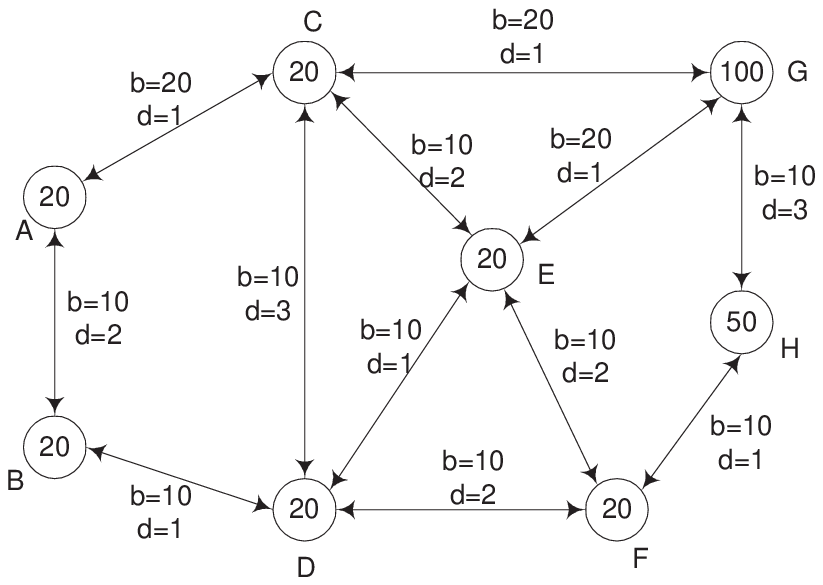}
\caption{An example resource network}
\label{fig:resource_graph}
\end{minipage}
\hspace{0.5cm}
\begin{minipage}{0.48\linewidth}[b]
\centering
\includegraphics{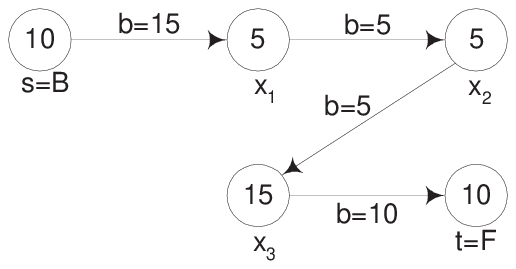}
\caption{An example stream processing task with a path topology}
\label{fig:problem_path}
\end{minipage}
\end{figure*}

Figure~\ref{fig:resource_graph} shows an example resource network of
eight interconnected computing nodes. Computational capacity of each
node is represented by a number inside the node. The link bandwidth
$(B)$ and costs $(d)$ are mentioned on each edge. An example stream
processing task of path topology with one source $s$, one sink $t$ and
three computational nodes $x_1$, $x_2$, $x_3$ is shown in
Figure~\ref{fig:problem_path}, with the node capacity and bandwidth
requirements. $s$ and $t$ must be mapped on $B$ and $F$,
respectively. There can be many feasible mappings of this dataflow
computation on the resource graph in
Figure~\ref{fig:resource_graph}. One of them is --

\begin{tabular}[h!]{c|c}
\begin{minipage}[t]{0.4\linewidth}
\begin{eqnarray*}
M(s) & = & B\\
M(x_1) & = & B\\
M(x_2) & = & B\\
M(x_3) & = & D\\
M(t)   & = & F\\
\end{eqnarray*}
\end{minipage}
&
\begin{minipage}[t]{0.48\linewidth}
\begin{eqnarray*}
M_e(s,x_1) & = & (B,B)\\
M_e(x_1,x_2) & = & (B,B)\\
M_e(x_2,x_3) & = & (B,D)\\
M_e(x_3,t) & = & (D,F)
\end{eqnarray*}
\end{minipage}
\end{tabular}

this is also the optimal solution in terms of total end-to-end cost
between the resource nodes $M(s)$ and $M(t)$.

We developed a decentralized algorithm that finds the exact solutions to
the problem. As the problem is NP-complete, some approximation scheme
is also proposed. The algorithm and approximation schemes are
discussed in Section~\ref{sec:mapping}.

\subsection{The Dynamic Re-allocation Problem}
\label{subsec:prob_dynasched}
Allocation of the server and link resources by the mapping process
would suffice, if all the resources were dedicated and under total
control of the platform. Because the data rate over the links through
the public network are variable and not under direct control of the
platform, a continuous adaptive allocation of the resources is
necessary. 

To minimize the overhead, it is desirable that the re-allocation is
done based on local information, otherwise a state-dissemination
protocol will be necessary. The global objective of re-allocation is
to maximize global processing throughput and keep the data-delivery
rate for each task as close as possible to the SLA defined delivery
rate. Locally, each server can monitor the rate at which it processes
data for each task using one of its services and the rate it transmits
data to the next service for each task. This information can be used
to determine how closely the task is progressing compared to the SLA
defined rate, because delivery rate is directly defined by the rate of
processing by each service component. Maximizing the compliance at each
server will imply maximum compliance at the delivery point. 

However, it is hard to know the global throughput of the processed
data from each server. We attempted to devise some local objectives,
achieving which would lead to achievement of the global
objective. Recall that each server allocates local resources
autonomously and also each server is paid for each byte of data it
processes. Rationally, each server would be inclined towards
maximizing its own revenue. We devised the allocation policy so that
it is consistent with such rational inclination of the servers. The
expectation is that maximization of the local revenue needs maximizing
local work throughput, which would lead to global throughput
maximization.

The adaptive reallocation is performed periodically at each server
node and the period need not be synchronized globally. In principle,
both the server and the link resources could be re-allocated. However,
in the proposed system model, servers are dedicated for stream
processing. A server accepts a task only if the requested amount of
processing resource is available. Thus, once a task gets server
resources allocated, its processing rate at that server does not vary
over time. However, transmission rate over the opportunistic network
links may vary over time, because they are shared resources and not
under complete control of the platform. Thus, to provide a predictable
performance guarantee for accepted tasks, it is essential to
adaptively re-allocate the link resources. On the other hand, although
re-allocation of the server resources could improve load balancing and
resource utilization because of changing load scenario, it is not
possible to re-map the task components on new servers locally or based
on local information, and the global mapping process has a lot of
overhead. For these reasons, we confined the re-allocation within
link resources only, leaving the initial allocation of server
resources unchanged.

The links that carry the stream between two data processing servers
can be of three different types -- i) a direct dedicated link, ii) a
multi-hop dedicated link through one or more forwarding nodes iii) an
overlay link through the public network. A mapping of a task may
contain any combination of these three types of links between the
processing nodes. Among them, the direct dedicated links are the most
preferred one, because they provide controlled and stable data rate. A
multi-hop dedicated link provides similar control and stability, but
it costs more (Section~\ref{subsec:model_pricing}).  The third
possibility is having an overlay link through the public network. The
flow rate is variable over such links, but there is no additional cost
for sending data through them. So, the nodes try to opportunistically
use these links when dedicated links are overloaded or not available.

The dynamic link scheduler in each node is invoked periodically at
regular intervals. Based on current evaluation of locally observed
performance, the scheduler re-allocates the locally available link
resources among the competing tasks that are using this node. The
overall policy of the scheduler is to prioritize the tasks based on
observed performance and re-allocate the three possible types of
outgoing links based on the newly estimated priorities. The
re-allocation process is illustrated in Figure~\ref{fig:dynasched}.
The re-allocation algorithm is described in
Section~\ref{sec:dynasched}, including the design of the appropriate
priority function.

 %talks about the mapping and scheduling problems
%%% Local Variables: 
%%% mode: latex
%%% TeX-master: "paper"
%%% End: 
\section{Algorithm for the Mapping Problem}
\label{sec:mapping}
In this section, we develop a decentralized algorithm for the
constrained path mapping problem introduced in
Section~\ref{subsec:prob_mapping}. The algorithm is then adapted
through some approximation heuristics and other modifications to use
in the bi-modal stream processing platform.

For distributed mapping, we use the scheme presented by Chandy and
Misra~\cite{Chandy1982}, which was based on Dijkstra and Scholten's
diffusing computation paradigm~\cite{Dijkstra1980}. The centralized
version of the problem, i.e. finding the mappings when the global
knowledge of the system state is available at a single node, can be
solved using the Bellman-Ford relaxation scheme. Such an algorithm was
analyzed in one of our previous works~\cite{HPCS2007}.

The distributed mapping algorithm uses two kinds of messages -- i)
{\em map} message and ii) {\em ack} message. The {\em map} messages
propagate the partially computed maps from the data-source node to the
data-delivery node through the network. The {\em ack} message is used
for detecting termination of the mapping algorithm, as commonly used in
diffusing computations. 

For each mapping, some variables are used to maintain the state of the
diffusing computation -- {\em count} maintains the number of
outstanding {\em ack} messages to be received against the sent {\em map}
messages.  {\em pred} maintains the name of the predecessor node in
the diffusing computation which made the current node aware of the
mapping by sending a {\em map} message when {\em count} was $0$.

To disseminate the task specification to all the participating nodes,
another type of message, the $spec$ message, is appended with the
$map$ message. To be efficient, when $u$ sends a map message to $v$,
it is sufficient to append the $spec$ with the $map$ only when $u$ is
the $pred$ for $v$. For this knowledge, every node maintains a flag
$knows(i)$ for each neighbor $i$, and sets the flag when a $spec$
appended $map$ message is sent to $i$. To assist this process, the
last {\em ack} sent by a node to its $pred$ when $count$ becomes $0$,
is differentiated from regular $ack$ using a $lastAck$ flag.  A node
resets the $knows(i)$ flag when it receives an $ack$ with the
$lastAck$ flag from its neighbor $i$. To save memory, each node
creates a state for the new mapping process only when it becomes aware
of the process by a $map$ message. The state is initialized ($pred =
undefined$, $count = 0$, $knows(i) = false$ for all $i$) on
creation. The state is removed when $count$ becomes $0$ and the
$lastAck$ is sent.

The mapping algorithm works primarily by enumerating all feasible
mappings on all possible paths. The optimal mapping is then chosen
from the feasible mappings. However, feasible mappings are gradually
expanded while exploring different paths and many of the mappings and
paths are pruned or discarded once any of the resource constraints
fails. Thus explicit enumeration of all possible alternatives are
avoided.

Each node executes the {\em processMap} method
(Algorithm~\ref{alg:mapping}) when a $map$ message is received and the
{\em processAck} method (Algorithm~\ref{alg:process_ack}) when an
$ack$ message is received.  Each time a node receives a partial map,
it tries to extend the partial map in all possible ways by appending
the mapping of more task components onto itself, subject to
availability of processing power~(Line~\ref{algmap:extension}).  Each
of these newly generated partial maps are then extended to all of the
neighbors as long as the bandwidth requirement of that hop in the task
is less than available bandwidth in that
link~(Line~\ref{algmap:sending}). Note that it is possible to extend
the map to the neighbors without having any component mapped on the
current node. This allows multi-hop connection between nodes
processing consecutive components. This is beneficial in cases where
there is no direct dedicated link between two server nodes. All the
feasible mappings are thus accumulated at the data-delivery node. The
acknowledgement process of the diffusing computation ensures
termination of the algorithm and allows each node to clear the states
related to the terminated mapping.

Cyclic mapping is allowed in the extension in
Line~\ref{algmap:sending}. Because $x=0$ is allowed, it is possible
that a mapping grows to an infinite length. In practice, this is
avoided by limiting the growth of the multi-hop mapping using a budget
factor. Based on the price-per-byte-processed quoted in the SLA
(Sections~\ref{subsec:model_task}~and~\ref{subsec:model_pricing}), the
allocated revenue for processing of the $j-th$ service is
limited. When the output of the $j$-th service is sent to the server
providing $(j+1)$th service using a dedicated link, host of the $j$-th
service needs to pay and thus loses revenue. The cost of transmission
grows as more dedicated links are used in a multi-hop link to send the
same data. Thus the number of hops in such multi-hop links are limited
by the revenue budgeted for the service and cost of each hop of
dedicated connection. This maximum hop restriction is summarized as
the {\em max\_null} parameter (Line~\ref{algmap:taskspec}) in the
Algorithm~\ref{alg:mapping}.

One point to note here is that the partial mappings cannot be pruned
using the optimality criterion, i.e.\ the cost metric. Even for the
same prefix of the task, a lower cost mapping may later get pruned by
the resource constraint while a higher cost mapping may survive. Thus
greedy pruning of the mappings based on the cost metric may not yield
the optimal solution. However, analysis in the
Section~\ref{subsec:mapping_heuristic} shows that such greedy pruning
dramatically reduces the number of messages without sacrificing too
much of the optimality.

%% present the algorithm
\begin{algorithm}[htbp]
    \caption{\em ProcessMap(m, T)}
    \label{alg:mapping}
    \begin{algorithmic}[1]
       \STATE {\label{line_input}{\bf Input:}} 

       \STATE { %% explain input variables 
         The current node executing the method is denoted as $v$. The
         sender of the message is $u$.}
       \STATE {\label{algmap:taskspec}
         $T = ${$t_1$, $t_2$, $\ldots$ , $t_{|T|}$} denotes the ordered set of 
         components in the stream processing task. Each $t_i$ has an
         associated $C(i)$ denoting processing
         requirement. Each $(t_i,t_{i+1})$ has an associated
         $B(i,i+1)$ denoting the required bandwidth. {\em max\_null}
         denotes the maximum number of empty hops allowed in the map. 
         $T$ is either found appended with the map message or from 
         the stored state.
       }
       \STATE{
         $m$ is the map message containing the mapping of the first 
	 $j$ services on a series of server nodes. $j$ is called the 
         {\em prefix-length} of $m$.  
       }
       \STATE{
	 For any node $u$, $C_{av}(u)$ denotes the computational
         capacity of $u$. $S(u)$ denotes the set of service components
         served by $u$. For a pair of nodes $u$ and $v$, $B_{av}(u,v)$
         denotes available bandwidth in the $(u,v)$ channel.
       }

       \IF{no state for $T$ or $pred$ is undefined}
          \STATE{store $T$ from the message}
          \STATE{create $pred$, $count$ and $knows$}
          \STATE{$pred \leftarrow u$, $count \leftarrow 0$,
          $\forall_{neighbor k, k \ne u}{knows(k)\leftarrow FALSE}$}
       \ELSE
          \STATE{Send $ack$(REGULAR) to $u$}
       \ENDIF

       \FOR{$x=0$ to $|T|-j-1$} 
             \IF {($x=0$) or ($t_{j+x} \in S(v)$ and 
                   $C_{av}(v) \geq \sum_{j=1}^x{C(j+x)} +
          \forall_{i\leq j}{t_i\mbox{ mapped on }v}\sum{C(i)}$)}
	        \STATE{\label{algmap:extension}$m_x \leftarrow $ 
                       map found by extending next 
                       $x$ services in $T$ on $v$}
		
		\IF{$v$ is the data-delivery node 
		  and ($j+x \geq |T|$)} 
		%% terminal node, feasible map found 
		   \STATE{store $m_x$ in the list of a feasible maps}
		\ENDIF
             \ELSE
                \STATE{break}
             \ENDIF
	     \FOR {each neighbor $k$ of $v$}
	        \IF {($B_{av}(v,k) \geq B(j+x, j+x+1)$)
                and (($x > 0$) or (empty hops in $m \leq
                max\_null-1$))} 
                \IF {$knows(k) = $FALSE}
                   \STATE{$knows(k) \leftarrow $ TRUE}
                   \STATE{Append $T$ to $m_x$}
                \ENDIF
	        \STATE{\label{algmap:sending}Send $m_x$ to $k$}
                \STATE{$count \leftarrow count + 1$}
		\ENDIF
	     \ENDFOR 
        \ENDFOR
    \end{algorithmic}
\end{algorithm}

\begin{algorithm}[htb]
\label{alg:process_ack}
\caption{processAck(isFinal)}
  \begin{algorithmic}
    \STATE{$ack$ message received from neighbor $u$}
    \STATE{$count \leftarrow count -1$}
    \IF{$count = 0$ and $pred$ is not invalid}
       \STATE{Send $ack$(FINAL) to $pred$}
       \STATE{$pred \leftarrow$ invalid}
    \ENDIF
    
    \IF {isFinal = FINAL}
       \STATE {$knows(u)\leftarrow$ FALSE}
    \ENDIF

  \end{algorithmic}
\end{algorithm}

\subsection{Heuristic Approximations}
\label{subsec:mapping_heuristic}
We observe that, in the worst case, the mapping algorithm in
Algorithm~\ref{alg:mapping} may generate all possible
source-destination paths in the graph and try all possible
combinations of the task components on each of those paths. Such
intractably explosive growth of complexity is expected because the path
mapping problem is NP-complete. For practical implementations, it may
be desirable to sacrifice some degree of optimality in favor of
reduction in the complexity. Here, we explore some heuristic
techniques that reduce the complexity while producing good
approximation for the optimal solution.

\subsubsection{LeastCostMap}
On intuitive way of reducing complexity is to greedily prune the
exploration of many of the alternative paths and mappings based on the
cost metric. In the {\em LeastCostMap} heuristic, a partial mapping
that has higher cost compared to a previously observed mapping of the
same prefix-length is pruned from further extension. To help this,
each node, for each task-mapping, maintains a table of the costs of
the least-cost partial mappings of each possible prefix lengths, among
the already observed partial mappings of the composite task. The cost
of the newly extended mapping in the Line~\ref{algmap:extension} of
Algorithm~\ref{alg:mapping} is compared to that in the table and is
sent to neighbors in Line~\ref{algmap:sending} only if the new mapping
has smaller cost. The cost in the table is updated accordingly.

\subsubsection{AnnealedLeastCostMap}
In the greedy pruning of higher cost partial maps, it is possible that
the mapping that would lead to the optimal solution is pruned while
the allowed mapping does not meet the constraints in the later
stages. One way to compromise between the greedy pruning and the
unpruned exponential growth of mappings is to apply a kind of
simulated annealing in the pruning process. A partial mapping of cost
higher than the already observed minimum is allowed for extension with
a probability and the probability diminishes exponentially with the
growing prefix-length of the mapping. This heuristic is hereafter
denoted as {\em AnnealedLeastCostMap} heuristic. Obviously, this
approach increases the message complexity, with the hope that some of
the non-minimal partial mappings would possibly lead to a better
complete mapping.

\subsubsection{RandomNeighbor}
Another way of restricting the message complexity is to extend a
partial map to a randomly chosen subset of $k$ neighbors instead of
expanding to all of them. Higher values of $k$ increases the chance of
getting the optimal solution. The {\em RandomNeighbor} heuristic with
$k=1$ did not produce results as good as LeastCostMap, although number
of messages were reduced dramatically. Further investigation may be
done to determine a suitable value of $k$.

\subsection{Performance of the Heuristics}
To choose one among the possible heuristics, we evaluated them running
the heuristics on an emulated network of nodes. We tried to measure
the quality of the approximate solutions generated by the heuristics
as well as their message overheads. The network topology was generated
by BRITE Internet topology generator~\cite{Brite2001}, using the
Barabasi-Albert algorithm~\cite{Barabasi1999}. This generates a
power-law graph and the link bandwidths were sampled from a truncated
power-law distribution having min=$10$Mbps and
max=$1$Gbps. Computational capacities of the nodes were randomly
assigned from a distribution of node-capacities of a volunteer
computing project~\cite{BOINCStatPaper2006}. The nodes were emulated
as processes hooked to UDP ports in LAN-connected computers. These
virtual nodes communicated among them using UDP packets. The network
size was varied from $30$ to $120$ nodes. The tasks for mapping
consisted of $10$ components. The bandwidth and capacity requirements
of each task-component was sampled from a Normal distribution with
mean equal to the $50\%$ of the average link and code capacity of the
network, respectively.

First, we attempted to evaluate how close the solutions generated by
the heuristics are to the exact optimal solutions. Because it is
computationally expensive to run the algorithm that gives the exact
optimal solution, we devised an algorithm that computes a lower bound
of the optimal solution. We relaxed the bandwidth constraints and
transformed the problem into finding a optimal cost path in a
multi-stage graph. The first and last stages resemble the source and
the terminal nodes.  Each of the internal stages have $n$ vertices,
resembling the choice of any of the $n$ servers for the processing
components of the tasks. Then we compute the lowest cost path from
source to the terminal vertex, subject to the node-capacity
constraints only. Ignoring the bandwidth constraints allows lower cost
solutions that are not feasible in the actual problem. All the
feasible solution for the actual problem will be feasible in the
relaxed problem. So, the optimal solution of the relaxed problem will
be a lower bound on the optimal cost of the actual problem. We
computed the ratio of the cost of heuristic generated solutions to
this lower bound cost. 

To assess the cost of executing the heuristics, we counted the total
number of {\em map} messages exchanged among the nodes. Because
arrival of each map message invokes the processing algorithm on the
receiving node, the total computational cost is also proportional to
the number of map messages. Although we did not evaluate the message
complexity of the exact algorithm, we have compared the complexities
of the heuristics, which helps to choose one heuristic over the
others.

\begin{figure}[htb]
  \centering
  \setcounter{subfigure}{0}
  \subfigure[Quality of solutions]{
    \includegraphics{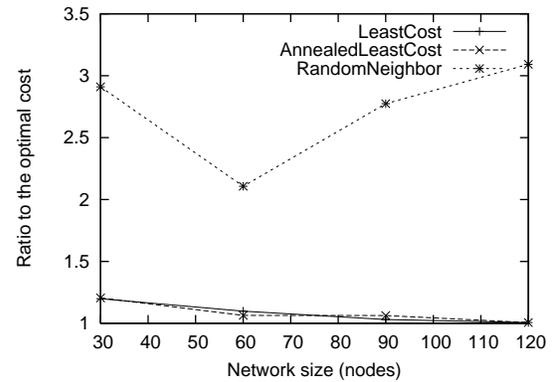}
    \label{fig:netmap_plot_optcost_graphsize}
  }
  \subfigure[Message overhead]{
    \includegraphics{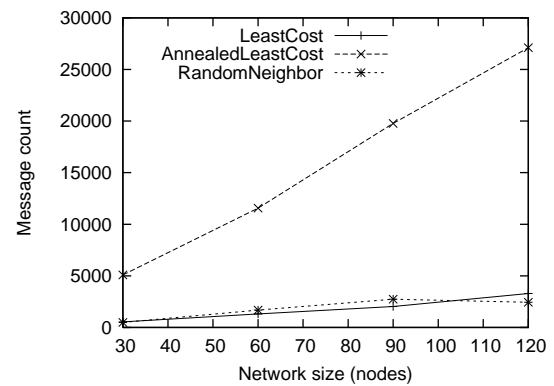}
    \label{fig:netmap_plot_msgcount_graphsize} 
  }
  \caption{Comparing three heuristics}
\end{figure}

Figure~\ref{fig:netmap_plot_optcost_graphsize} shows that the
heuristic derived solutions are fairly close to the lower bound of the
optimal solutions. One can observe that both the {\em LeastCostMap}
and the {\em AnnealedLeastCostMap} yield solutions that are equally
very close to the optimal solutions. The {\em RandomNeighbor}
heuristic does not produce good solutions, because number of feasible
ways to expand the partial maps narrows down very quickly here. In
terms of cost of computation of the heuristics, we observe in
Figure~\ref{fig:netmap_plot_msgcount_graphsize} that number of {\em
map} messages to complete mapping of a single task-composition is much
higher in the {\em AnnealedLeastCostMap} heuristic than the other two
heuristics. This is because, for the chosen parameter setting, the
{\em AnnealedLeastCostMap} extends many more of the alternative paths
and mappings compared to the {\em LeastCostMap} heuristic. Analyzing
both the results in Figure~\ref{fig:netmap_plot_optcost_graphsize} and
Figure~\ref{fig:netmap_plot_msgcount_graphsize}, it may be concluded
that the additional message overhead due to the late pruning in the
{\em AnnealedLeastCostMap} does not worth its gain in optimality.
Finally, we chose the {\em LeastCostMap} heuristic for our distributed
stream processing platform.

\subsection{Modifications for Bi-Modal Communication  Links}
So far, in the design of the decentralized mapping algorithm, we did
not consider the presence of the opportunistic communication links. As
mentioned in the system model in Section~\ref{sec:model}, each node is
connected to the public Internet and can establish an end-to-end
connection with any other node. The presence of these all-to-all links
require some modifications in the {\em ProcessMap} procedure described
in Algorithm~\ref{alg:mapping}.

Because, with opportunistic links, all other nodes in the platform are
neighbors in terms of connectivity, sending of extended maps to all
neighbors in Line~\ref{algmap:sending} of {\em ProcessMap} would be
inefficient, although it would work. Instead, after extending the
mappings to all the dedicated link neighbors, the mappings may be
extended to a small subset of the opportunistic-link-neighbors. To
choose a subset, we assume that each node has an approximate knowledge
of which node serves which service. We assume that there exist a
gossip mechanism to disseminate this knowledge. Note that the set of
services available at a node changes much less frequently compared to
arrival of individual task mapping requests. Moreover, this knowledge
is used only for minimizing the overhead, thus its inaccuracy does not
harm much other than missing some possible solutions. Another point to
note is that having all-to-all connectivity, there is no meaning of
mapping a hop of the task-composition on multiple hops of
opportunistic links, although multi-hop dedicated connection is still
preferable.

\begin{algorithm}[htbp]
    \caption{\em ProcessMap2(u, m, T)}
    \label{alg:mapping2}
    \begin{algorithmic}[1]
       \STATE {\label{line_input}{\bf Input:} As described in
       Algorithm~\ref{alg:mapping}}

       \IF{no state for $T$ or $pred$ is undefined}
          \STATE{store $T$ from the message}
          \STATE{create $pred$, $count$ and $knows$}
	  \STATE{\label{algmap2:M}create $M(1:|T|)$} 
          \STATE{$pred \leftarrow u$, $count \leftarrow 0$,
          $\forall_{neighbor k, k \ne u}{knows(k)\leftarrow FALSE}$}
	  \STATE{$\forall_i{M(i) \leftarrow \inf}$}
       \ELSE
          \STATE{Send $ack$(REGULAR) to $u$}
       \ENDIF

       \FOR{$x=0$ to $|T|-j-1$} 
             \IF {($x=0$) or ($t_{j+x} \in S(v)$ and 
                   $C_{av}(v) \geq \sum_{j=1}^x{C(j+x)} +
          \forall_{i\leq j}{t_i\mbox{ mapped on }v}\sum{C(i)}$)}
	        \STATE{\label{algmap2:extension}$m_x \leftarrow $ 
                       map found by extending next 
                       $x$ services in $T$ on $v$}
		
		\IF{$v$ is the data-delivery node 
		  and ($j+x \geq |T|$)} 
		%% terminal node, feasible map found 
		   \STATE{store $m_x$ in the list of a feasible maps}
		\ENDIF
             \ELSE
                \STATE{break}
             \ENDIF

	     \IF{\label{algmap2:Mcondition}$cost(m_x) < M(|m_x|))$}
	        \FOR {each dedicated-link-neighbor $k$ of $v$}
	          \IF {($B_{av}(v,k) \geq B(j+x, j+x+1)$)
                  and (($x > 0$) or (empty hops in $m \leq
                  max\_null-1$))} 
                    \IF {$knows(k) = $FALSE}
                       \STATE{$knows(k) \leftarrow $ TRUE}
                       \STATE{Append $T$ to $m_x$}
                    \ENDIF
	            \STATE{\label{algmap2:sending_ded}Send $m_x$ to $k$}
                    \STATE{$count \leftarrow count + 1$} 
		  \ENDIF
	        \ENDFOR 
		\IF {\label{algmap2:opp_loop_start}$x > 0$} 
                  \FOR {each node $k$ such that $k$ provide 
                        the service $j+x+1$} 
                     \IF {available uplink bandwidth to the Internet
                        $\geq$ bandwidth need for service hop $(j+x, j+x+1)$} 
                        \STATE{\label{algmap2:sending_opp}Send $m_x$ to $k$}
                     \ENDIF
                  \ENDFOR 
		\ENDIF {\label{algmap2:opp_loop_end}}
	     \ENDIF % cost (m_x)
        \ENDFOR
   \end{algorithmic}
\end{algorithm}

The final version of the {\em ProcessMap} algorithm that applies the
{\em Least CostMap} heuristic and takes care of opportunistic links is
presented in Algorithm~\ref{alg:mapping2}. The $M(1:|T|)$ data
structure (Line~\ref{algmap2:M}) to store the costs of the
minimum-cost mapping among the already observed partial maps, and the
condition in Line~\ref{algmap2:Mcondition}, are added for the {\em
  LeastCostMap} heuristic. The other additional code in
Lines~\ref{algmap2:opp_loop_start}-\ref{algmap2:opp_loop_end} handles
the extension of the mappings through opportunistic links. Note that
such extension is allowed only when at least one task-component is
mapped on the current node
(Line~\ref{algmap2:opp_loop_start}). Because it is not possible to
allocate end-to-end bandwidth in the opportunistic links, only the
uplink bandwidth is allocated. The end-to-end bandwidth that a task
actually gets is monitored and reactively allocated in a continuous
feedback loop, which we will discuss in the next section.

To devise an appropriate cost metric for choosing the best among
alternative feasible maps, we considered the following two factors -
balancing the service workload among the servers and minimizing the
uncertainty of using opportunistic links.  The load-balance factor
for a map (or a partial map) is computed as an average of the server
load-factors (ratio of used capacity to total capacity) for all the
servers included in the map, and is always a number between $0$ and
$1$. A map with lower load-balance factor spreads the components of a
task on different servers rather than putting all of them into one,
and chooses the under-utilized servers.  In case two maps have almost
same load-balance factor, (do not differ by more than $0.1$ or
$10\%$), then the one in which the number of hops (links connecting
the processing components) assigned to dedicated links is higher is
preferred. If that is also same, the map with least number of hops
through public network is preferred.

 %solving the mapping problem, evaluation of the
		%heuristics may go here
\section{Adaptive Re-allocation of the Bi-modal Links}
\label{sec:dynasched}
The dynamic link scheduler in each node is invoked periodically at
regular intervals. Based on current evaluation of locally observed
performance, the scheduler re-allocates the locally available link
resources among the competing tasks that are using this node. The
overall policy of the scheduler is to prioritize the tasks for use of
the network links, based on their deviation from target data rate and
the price they would pay for the data processing service.

The links that carry the stream between two data processing servers
can be of three different types -- i) a direct dedicated link, ii) a
multi-hop dedicated link through one or more forwarding nodes iii) an
overlay link through the public network. A mapping of a task may
contain any combination of these three types of links between the
processing nodes. Among them, the direct dedicated links are the most
preferred one, because they provide controlled and stable data rate. A
multi-hop dedicated link provides similar control and stability, but
it costs more (Section~\ref{subsec:model_pricing}).  The third
possibility is having an overlay link through the public network. The
flow rate is variable over such links, but there is no additional cost
for sending data through them. So, the nodes try to opportunistically
use these links when dedicated links are overloaded or not available.

\begin{algorithm}[htbp]
    \begin{algorithmic}[1]
      \STATE{Invoked for each node $u$ periodically}
      \STATE{\label{algsched:group}Group the tasks that are being processed in $u$ by their
      next hop server $v$}
      \FOR{\label{algsched:priority_start}Each group $v$}
           \STATE{Compute the priority of each flow competing for a
      ($u$,$v$) link as -}
	   \STATE{priority $\leftarrow$ budget per byte of processed data *
      bandwidth required to comply with the target rate}
	   \IF {any dedicated link ($u$,$v$) exists}
	   \STATE{\label{algsched:ded_assign}Assign the dedicated link to top priority flows
      until all capacity is used}
	   \ENDIF
	   \STATE{Collect all the unassigned flows} 
      \ENDFOR \label{algsched:priority_end}
      \FOR{ All the remaining flows}
       \IF{The budget permits $k$-hop ($u$,$v$) dedicated link, $k>1$}
       \STATE{\label{algsched:multihop_assign}Launch a probe search 
            and reserve multi-hop dedicated
            path for the flow with maximum $k$ hops}
       \STATE{Assign public network bandwidth for the flow temporarily}
       \ELSE
       \STATE{\label{algsched:public_assign}Assign public network 
             bandwidth for the flow}
       \ENDIF
      \ENDFOR
   \end{algorithmic}  
   \caption{Link re-allocation algorithm}
   \label{alg:netdyna_schedalgo}
\end{algorithm}

Algorithm~\ref{alg:netdyna_schedalgo} is executed when the scheduler
is invoked at regular intervals.  The algorithm evaluates the For
allocation of the links, tasks are grouped according to their next hop
server node (Line~\ref{algsched:group}). While prioritizing among
competing tasks for each group
(Lines~\ref{algsched:priority_start}-\ref{algsched:priority_end}), the
scheduler tries to maximize the revenue earning of the server and
prefers the tasks marked with higher price per unit of processing. On
the other hand, the server tries to fulfill the rate requirement of
each task, because it gets penalized otherwise. Hence the scheduler
computes the priority of each task as a product of the apportioned
price and the data rate required in next scheduling epoch.

For each next hop group, highest priority tasks get allocation from
the direct dedicated link, if such a link exists and capacity permits
(Line~\ref{algsched:ded_assign}). The next prior tasks are assigned
multi-hop dedicated links (Lines~\ref{algsched:multihop_assign}). The
maximum possible hops in such multi-hop links are restricted by the
apportioned price for that service according to the task
specification.  The remaining tasks from all the groups are allocated
bandwidth from the opportunistic public network links
(Line~\ref{algsched:public_assign}).

\subsection{Is Adaptive Re-Allocation Necessary?} 
So far, we have argued that due to the inconsistent behavior of the
opportunistic links, it is necessary to reallocate the link resources
periodically in a feed-back loop. Here we asses the necessity of such
re-allocation quantitatively. 

The main intuition behind introducing dynamic re-allocation is that
the data-stream that goes through the public network suffers from the
variability and lag from the target rate, whereas the stream that uses
dedicated links all-through, does not lag from the target at
all. Dynamic scheduling introduces fairness across all the tasks.  So
if link assignment is done dynamically, it is expected to improve the
utilization of the resources and increase the overall work-throughput
of the system.

\begin{figure*}[htb]
  \centering
  \setcounter{subfigure}{0}
    \subfigure[Throughput]{
          \label{fig:netdyna_throughput_arvrate_comparesched}
          \includegraphics[scale=0.98]{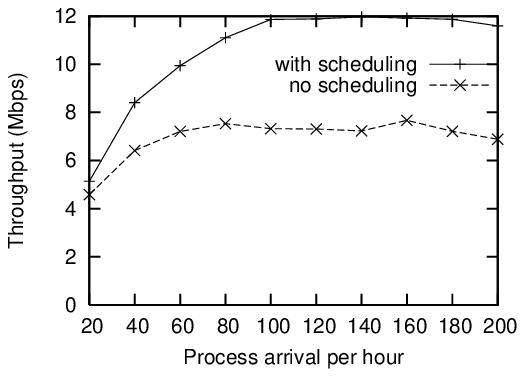}
    }
    \subfigure[Utilization of dedicated links]{
          \label{fig:netdyna_dedutilnet_arvrate_comparesched}
          \includegraphics[scale=0.98]{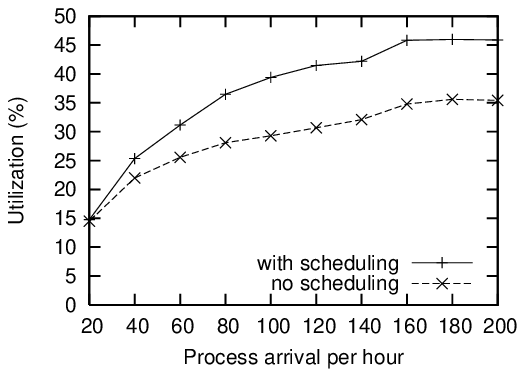}
    }
    \subfigure[Deviation from target rate]{
          \label{fig:netdyna_deviation_arvrate_comparesched}
          \includegraphics[scale=0.98]{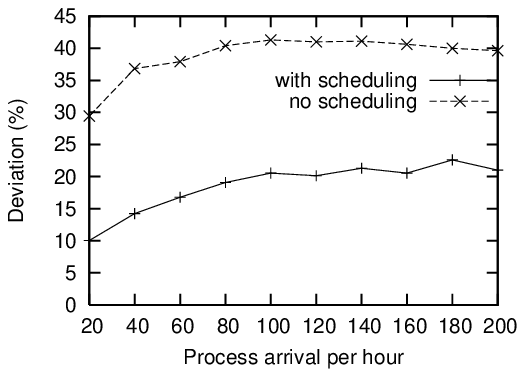}
    }
    \caption{Effect of dynamic scheduling}
\end{figure*}

For the evaluation we used a $100$-node simulated stream processing
platform. Details of the simulation set-up is described later in
Section~\ref{subsec:netdyna_sim_model}. We fed the same workload to
two system set-ups, both having bi-modal communication networks. In
one, we disabled the adaptive re-allocation of links and let the tasks
complete with the initial assignment of links and nodes. The adaptive
re-allocation is enabled in the other. All other system parameters
were the same for both the set-ups. From
Figures~\ref{fig:netdyna_throughput_arvrate_comparesched} we observe
that overall system throughput increases with adaptive re-allocation,
as an indication of higher task acceptance ratio and higher
utilization of the system resources.
Figure~\ref{fig:netdyna_dedutilnet_arvrate_comparesched} demonstrates
that adaptive re-allocation results in much higher utilization of the
dedicated links. CPU utilization remains unchanged (not shown),
because the dynamic re-allocation does not alter the node
assignments. Another rationale behind re-allocations is to increase
fairness and improve compliance with the target delivery
rate. Figure~\ref{fig:netdyna_deviation_arvrate_comparesched} shows
that irrespective of workload, the adaptive re-allocation decreases
the deviation from the specified target rate.

 %solving the dynamic scheduling problem,
		     %evaluation regarding the presence and absence of
		     %scheduling may go here
\section{Performance Evaluation and Discussion}
\label{sec:results}
\subsection{Simulation Model}
\label{subsec:netdyna_sim_model}
We constructed a simulation model of the distributed stream processing
platform according to the architecture and algorithms presented in
Sections~\ref{sec:model} and~\ref{sec:problem}, respectively. The
model was build on Java based simulation engine JiST~\cite{Jist2005}. 

Each of the servers in distributed locations are connected to the
public Internet. Although each server has a certain uplink and
downlink bandwidth, the data rate over a connection that goes through
the public network faces temporal variation.  We use the statistics
presented by Wallerich and Feldmann~\cite{Wallerich2006} to model the
temporal variability of the end-to-end capacity of a path through the
public network. From their data collected from packet level traces
from core routers of two major ISPs over 24 hours, the logarithm of
the ratio of the observed transient flow rate to the mean flow rate
over long period is almost a Normal distribution. In our simulations,
all flows on the public network are perturbed every $10$ milliseconds
according to this model. With the allocated bandwidth as the mean rate
and the standard deviation of the log-ratio set at $1$, in $95\%$ of
the cases the observed bandwidth remains between one fourth
($2^{-2\sigma}$) and four time ($2^{2\sigma}$) of the allocated or
mean bandwidth. Bandwidth of each last-mile connection (uplink and
downlink) is randomly assigned between $1$ Mbps and $2$ Mbps.

In addition to the public network links, the servers are
interconnected through dedicated links (which may be leased lines or
privately installed links). For the dedicated network, we assume a
preferential connectivity based network growth model similar to the one
proposed by Barabasi et al~\cite{Barabasi1999}. The basic premise here
is that when a server attempts to establish a dedicated link, it does
so preferably with the most connected server. This eventually results
in a power law degree distribution in the network. We assumed that
server CPU capacity is proportional to the number of dedicated links
it has. The variety of services that a server can host is also
proportional to the node degree or capacity.  The dedicated links have
much higher bandwidth than the network links connecting a node to the
public network. Their bandwidths were randomly assigned between $1$
Mbps and $10$ Mbps and the propagation delays were assumed to be
between $1$ and $10$ milliseconds.  The propagation delay of an
end-to-end connection through the public network was much higher and
assumed to be between $10$ and $100$ milliseconds.

\begin{figure*}[htb]
  \centering
  \setcounter{subfigure}{0}
  \subfigure[Task Throughput]{
    \label{fig:netdyna_throughput_numlinks}
    \includegraphics[scale=0.98]{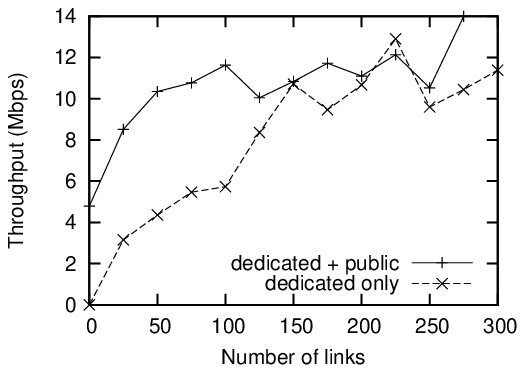}
  }
%  \hspace{.1cm}
  \subfigure[Task acceptance ratio]{
    \label{fig:netdyna_acceptratio_numlinks}
    \includegraphics[scale=0.98]{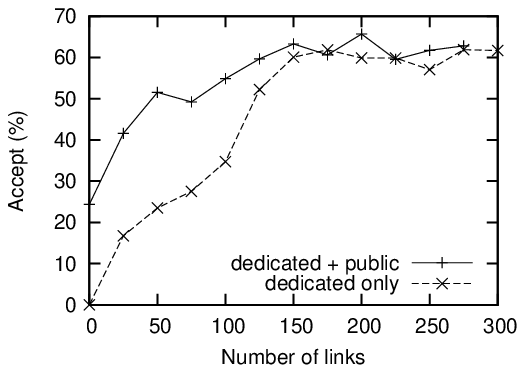}
  }
%  \hspace{.1cm}
  \subfigure[Server utilization at different workload]{
    \label{fig:netdyna_cpuutilnet_arvrate}
    \includegraphics[scale=0.98]{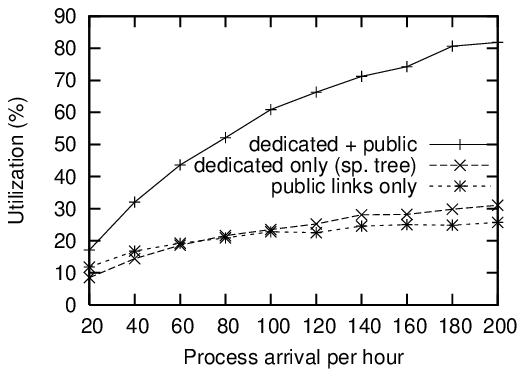}
  }
  \vspace{.1cm}
  \subfigure[Server utilization vs dedicated links]{
    \label{fig:netdyna_cpuutilnet_numlinks}
    \includegraphics[scale=0.98]{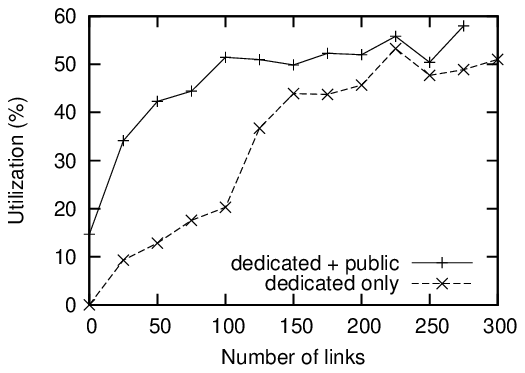}
  }
%  \hspace{.1cm}
  \subfigure[Link utilization at different workload]{
    \label{fig:netdyna_dedutilnet_arvrate}
    \includegraphics[scale=0.98]{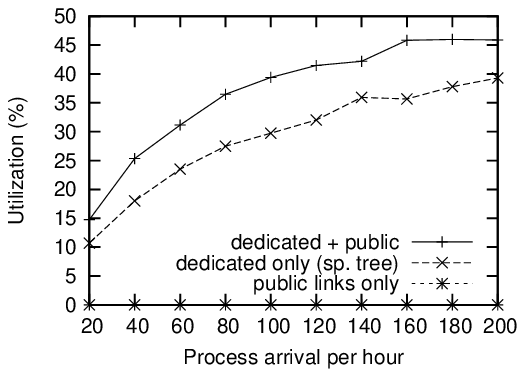}
  }
%  \hspace{.1cm}
  \subfigure[Link utilization vs number of dedicated links]{
    \label{fig:netdyna_dedutilnet_numlinks}
    \includegraphics[scale=0.98]{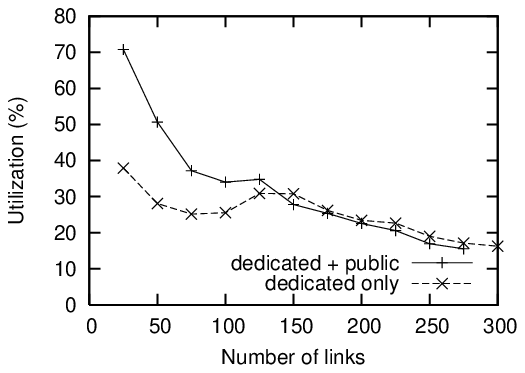}
  }
  \vspace{.1cm}
  \subfigure[SLA deviation at different workload]{
    \label{fig:netdyna_deviation_arvrate}
    \includegraphics[scale=0.98]{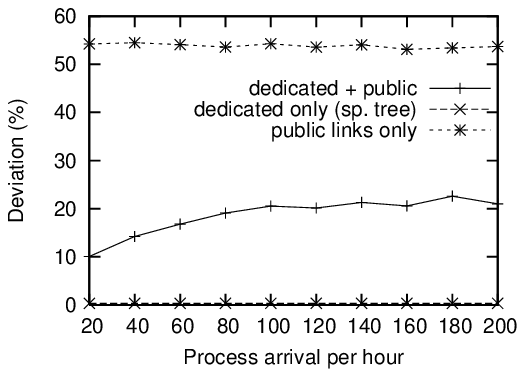}
  }
%  \hspace{.1cm}
  \subfigure[SLA deviation vs number of dedicated links]{
    \label{fig:netdyna_deviation_numlinks}
    \includegraphics[scale=0.98]{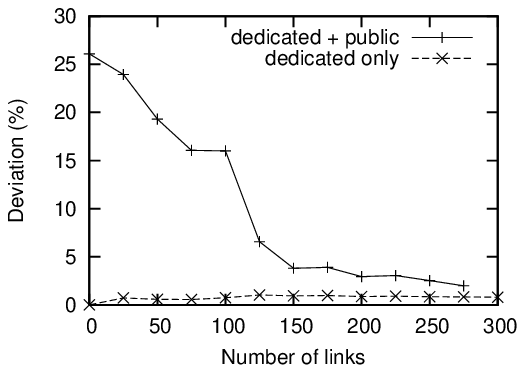}
  }
%  \hspace{.1cm}
  \subfigure[Elongation of task execution time]{
    \label{fig:netdyna_exectime_arvrate}
    \includegraphics[scale=0.98]{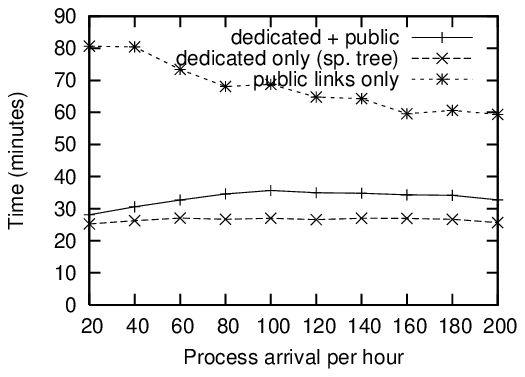}
  }
  \caption{Comparing bi-modal and uni-modal networks}
\end{figure*}

Unless otherwise mentioned, we assumed the platform to have $100$
server nodes and $99$ dedicated links interconnecting them. There were
$25$ different types of services. As the service variety is
proportional to the node degree, a node having $d$ dedicated links was
assumed to host $1+d$ different types of services (one added for
public network link). Server CPU capacity was set such that it can
execute $k$ instances of each service concurrently, according to the
mean data delivery rate. We set $k=2$. For the task workload, each
task is assumed to have $10$ service components, randomly chosen from
$25$ different types of service. Mean data delivery rate was $1$Mbps
and total amount of data to be processed from the source was $100$MB
on average. Each data point on the results shown below is an average
of $100$ observations from different experiments on randomly generated
networks with specified parameters. For each experiment, a synthetic
workload trace containing $500$ stream processing tasks were
generated. The task arrival process is assumed to be Poisson, with the
arrival rate varying across the experiments. If not mentioned
otherwise, the default arrival rate was $60$ tasks per hour.

\subsection{Benefits of Combining Opportunistic and Dedicated Resources}
We performed several sets of experiments to evaluate the benefits of
using bi-modal networks for stream processing tasks.
In  the experiments, we compare three possible settings
-- i) a network with the dedicated links only, ii) public network
only, and iii) a network that combines both. 

First argument in favor of a bi-modal network for stream processing is
that combining the public network with dedicated links, the system
achieves much higher work throughput at the same cost. To examine
this, we fed similar workload traces under same arrival rates to two
system set-ups, one with only dedicate link based networks and the
other using the combination of dedicated links and public
network. From Figure~\ref{fig:netdyna_acceptratio_numlinks} we observe
that for the same workload, if the platform uses dedicated links only,
it needs more than $120$ links to get $50\%$ acceptance ratio, whereas
the same acceptance ratio can be obtained with $50$ dedicated links
only, if the public network is utilized in conjunction. Similar
evidence in Figure~\ref{fig:netdyna_throughput_numlinks} shows that
inclusion of the public network helps to achieve same overall system
throughput at much lower number of dedicated link installations.

The next argument is that utilization of the privately deployed
expensive dedicated resources such as servers and dedicated links is
increased, if inexpensive public network is used in conjunction.  From
Figure~\ref{fig:netdyna_cpuutilnet_arvrate} we observe that when a
combination of dedicated links and the public network is used, the server
utilization is higher than the sum of utilizations of 
cases using a single type of network links.

Figures~\ref{fig:netdyna_dedutilnet_arvrate}
and~\ref{fig:netdyna_dedutilnet_numlinks} show another evidence of
higher return on investment. In
Figure~\ref{fig:netdyna_dedutilnet_arvrate}, we observe that the
utilization of dedicated links becomes consistently higher across a
wide range of loading scenarios if the public network is used in
combination. The lower utilization in case of a dedicated link only
network results from the fact that the platform has rejected many task
requests that would have been feasible by the augmentation of the
public resources. Figure~\ref{fig:netdyna_dedutilnet_numlinks} shows
the variation of utilization of the dedicated links with the number of
dedicated links. We observe that the difference in utilization
diminishes as the number of installed links increases. This is because
when there is sufficient number of dedicated links to carry the
required traffic of all the tasks, the public resources are not used
at all, and the bi-modal system becomes equivalent to a dedicated link
only system. In both cases, utilization of the links keeps decreasing
when more and more links are added because the workload is held
constant.

The discussion above highlighted the benefits of using public network
towards improving the utilization of dedicated server and link
resources (i.e., increases in return on investment).  Next we
investigate how the bi-modal network helps the stream processing
platform to improve the compliance with the services contracts it has
with individual tasks. We measure the compliance of the stream
processing platform as follows.  Each task request specifies a time
window $T$ that is used to monitor the delivery rate. We measured the
deviation from the required rate as $\sum_{\mbox{over all
windows}}\frac{B - \hat{B}}{B}$, where $B$ is the desired rate and
$\hat{B}$ is the observed rate of delivery. In
Figure~\ref{fig:netdyna_deviation_arvrate}, we observe that use of
dedicated links brings the percent deviation down to between $10\%$
and $20\%$ from above $50\%$. In this case the number of installed
dedicated links was just enough to make a spanning tree of the nodes,
i.e.\ $N-1$ links for $N$ nodes. Note that deviation is counted on the
accepted jobs only. So, even though for a dedicated link only network,
the deviation is almost zero, we have seen that such network is unable
to accept enough jobs to fully utilize the resources. In
Figure~\ref{fig:netdyna_deviation_numlinks}, we observe that the
deviation in the bi-modal system gets closer to zero as more and more
dedicated links are added to the network. However, beyond certain
number of links, ($125$ in this particular experiment), the
improvement is very marginal.

When we use a combination of dedicated and public links, it is
expected that the completion time of each task will be slightly
elongated compared to a system with only dedicated links, due to the
variability in the public network. Nevertheless, using the combination
contains the elongation to a small value, compared to the case where
only public network is available. In
Figure~\ref{fig:netdyna_exectime_arvrate}, we observe a $10-20\%$
increase in the execution time in the bi-modal system, whereas
execution time would be $200-300\%$ more in case of a public network
only system.

 %overall performance evaluation
\section{Related Work}
\label{sec:related}

Although there is a vast body of literature on resource management in
cluster, Grid or peer-to-peer hosting platforms, there have been
relatively a very few works that proposes combined use of dedicated
and public resources. In~\cite{HSQ2003}, Kenyon et al.\ provided
arguments based on mathematical analysis, that commercially valuable
quality assured services can be generated from harvested public
computing resources, if small amount of dedicated computers can
be augmented with them. With simple models for available periods
of harvested cycles, their work have measured the amount of dedicated
resources necessary to achieve some stochastic quality assurance from
the platform. However, they did not study how a bi-modal platform
would perform in the presence of clients with different service level agreements
and how to engineer the scheduling policies to maximize
the adherence to these agreements.

Recently, in~\cite{Kleinrock2006}, Das et al.\ have proposed the use
of dedicated streaming servers along with BitTorrent, to provide
streaming services with commercially valuable quality assurances while
maintaining the self scaling property of BitTorrent platform. With
analytical models of BitTorrent and dedicated content servers they
have demonstrated how guaranteed download time can be achieved through
augmentation of these platforms. However, their proposal does not
include actual protocols that can be used to achieve these performance
improvements.

Architectures and resource management schemes for distributed stream
processing platforms have been studied by many research groups from
distributed databases, sensor networks, and multimedia
streaming~\cite{Repantis2009, Benoit2009, Pietzuch2008, Hwang2008}.
In database and sensor network research, the major focus was placing
the query operators to nodes inside the network that carries the data
stream from source to the viewer~\cite{Pietzuch2006}. In multimedia
streaming problems, similar requirements arise when we need to perform
a series of on-line operations such as trans-coding or embedding on
one or more multimedia streams and these services are provided by
servers in distributed locations. In both cases, the main problem is
to allocate the node resources where certain processing need to be
performed along with the network bandwidths that will carry the data
stream through these nodes.

Finding the optimal solution to this resource allocation problem is
inherently complex. Several heuristics have been proposed in the
literature to obtain near-optimal solutions.
Recursive partitioning of the network of computing nodes have been
proposed in~\cite{Schwan2005} and~\cite{Seshadri2007} to map the
stream processing operators on a hierarchy of node-groups. They have
demonstrated that such distributed allocation of resources for the
query operators provides better response time and better tolerance to
network perturbations compared to planning the mapping at a
centralized location.

In~\cite{Baochun2004} and~\cite{Gu2006}, the service requirements for
multi-step processing of multimedia streams, defined in terms of
service composition graphs have been mapped to an overlay network of
servers after pruning the whole resource network into a subset of
compatible resources. The mapping is performed subject to some
end-to-end quality constraints, but the CPU requirements for each
individual service component is not considered.
Liang and Nahrstedt in~\cite{Nahrstedt2006} have proposed solutions to 
the mapping problem where both node capacity requirement and bandwidth
requirements are fulfilled. However, one of the assumptions made by
Liang and Nahrstedt was that the optimization algorithm was executed
in a single node and complete state of the resource network is
available to that node before execution. In a large scale dynamic
network this assumption is hard to realize. If we assume that each
node in the resource network is aware of the state of its immediate
neighborhood only, we need to compute the solution using a distributed
algorithm such as ours.

In all of the abovementioned works, the operator nodes are assumed to
interconnected through an application dependent overlay network using the
Internet as underlay. In~\cite{Nahrstedt2003}, Gu and Nahrstedt
presented a service overlay network for multimedia stream processing,
where they have shown that dynamic re-allocation of the operator nodes
provides better compliance with the service contracts in terms of
service availability and response time. However, none of the works
have proposed the use of dedicated links in conjunction with IP
overlay network for improving adherence to the service contracts. %related work 
\section{Conclusion}
\label{sec:conc}

In this paper, we investigated the resource management problem with
regard to data stream processing tasks. In particular, we examined how
a hybrid platform made up of dedicated server resources and bi-modal
network resources (dedicated plus public) can be used for this class
of applications. From the simulation based investigations, we were
able make several interesting observations. First, bi-modal networks
can improve dedicated resource utilization (server plus dedicated
network links). This means higher return on investment can be obtained
by engaging the bi-modal network. Second, the overall system is able
to admit and process tasks at a higher rate compared to system
configurations that do not leverage a bi-modal network. Because the
public network is engaged at zero or very low cost, this improvement
in throughput can be result in significant economic gain for
institutions that perform data stream processing workloads. Third, the
engagement of bi-modal network comes at a slight overhead that adds
low delays in stream processing tasks. Compared to public-only
networks the delays provided by the bi-modal network is almost
negligible. Fourth, dynamic rescheduling is essential to cope with
varying network conditions -- particularly in the public network. The
dynamic rescheduling algorithm switches the flows according to the
recomputed priority values to achieve the best service level
compliances.

In summary, our study highlights the benefits of the bi-modal
architecture for compute- and network-intensive
applications. Moreover, it provides simple distributed algorithms that
allows the effective utilization of such a platform for data stream
processing applications. Deploying the distributed resource management
framework in an actual prototype for data stream mapping is a possible
future work.

%\singlespace
%End----------------------------------------------------------------------
%
\balance
\bibliographystyle{IEEEtran}
\bibliography{asad}

\end{document}